\documentstyle[twoside,fleqn,espcrc2,epsf,rotate]{article}

\title{Glueball matrix elements on anisotropic lattices}

\author{Kentucky Glueball Collaboration:
        S.J. Dong,\address{Department of Physics and Astronomy,
                  University of Kentucky,
                  Lexington, KY 40506, USA.}
        T. Draper$^{\rm a}$,
        W.C. Kuo,\address{Department of Physics,
                 National Sun Yat-Sen University,
                 Kaohsiung 804, Taiwan.}
        K.F. Liu$^{\rm a}$,
        C. Nenkov$^{\rm a}$\thanks{Presented by C.\ Nenkov at Lattice '97, 
                                   Edinburgh -- Scotland},
        \\M. Peardon$^{\rm a}$,
        J. Sloan$^{\rm a}$,
        and B.L. Young\address{Department of Physics and Astronomy,
                      Iowa State University,
                      Ames, IA 50011, USA.}
        \thanks{This work is supported in part by the U.S. Department
                of Energy under grant numbers DE-FG05-84ER40154 and 
                DE-FC02-91ER75661, and by the Center for
                Computational Sciences, University of Kentucky.}}

\begin{document}

\begin{abstract}

We describe a lattice calculation of the matrix elements relevant for
glueball production in $J / \psi$ radiative decays.  The techniques
for such a calculation on anisotropic lattices with an improved
action are outlined.  We present preliminary results showing the
efficacy of the computational method.

\end{abstract}

\maketitle

\section{MOTIVATION}\label{sec:motive}

Glueballs, apart from being one of the most exotic states predicted
from QCD, are crucial in completing the picture of the strong
interaction, since these mesonic states go beyond the standard quark
model.  Unfortunately there is a considerable controversy as to their
phenomenological identification.  Even for the low-lying states there
are a lot of experimental candidates:
\begin{itemize}
\item \[
0^{++} -\!\!\!- \left\{ \begin{array}{l}
  \mbox{$f_{0}(1500)$ from $p \bar{p}$ annihilation,} \\ 
  \mbox{$f_{J}(1710)$ from radiative $J/ \psi$} \\
  \mbox{\ \ \ \ \ \ \ \ \ \ \ \ \ decay} 
  \end{array} \right. \]
\item
$0^{-+}$ --- $\eta (2225)$ 
\item \[
2^{++} -\!\!\!- \left\{ \begin{array}{l}
  \mbox{$f_{2}(1710)$ and $f_{2}(2230)$ from} \\
  \mbox{radiative $J/ \psi$ decay} 
  \end{array} \right. \]
\end{itemize}

Recently, there appears to be strong experimental evidence that
glueballs have been detected~\cite{glue_expt}.  For years radiative
$J / \psi$ decays,
\begin{equation}
J / \psi\ -\!\!\!-\!\!\!-\!\!\!\rightarrow\ \gamma\ +\ G 
\end{equation} 
have been considered \cite{liu89} the best hunting ground for low
lying glueballs (i.e.\ those below 3 GeV).

The main objective of this work is to measure vacuum-to-glueball
transition matrix elements which are needed for estimating the
partial widths of $J / \psi$ radiative decays into glueballs
\cite{liu93}.  We employ an $O(a_s^2)$-improved $SU(3)$ gauge action
on an anisotropic lattice \cite{action}, with the temporal lattice
spacing much less than the spatial one.  This has been
shown~\cite{pm1} to be an efficient means of studying the masses of
glueballs, as clearer resolution of the decay of correlators is
allowed.  The use of this technique in a matrix element calculation
seems highly advantageous as the local current operator introduces
more UV noise into Monte-Carlo measurement.

\section{METHOD}\label{sec:method}

The matrix elements we compute are of the form $\langle
G|F^{a}_{\mu\nu}F^{a}_{\rho\sigma}|0 \rangle$, where $G$ is the
$0^{++},0^{-+}$ or $2^{++}$ glueball and $F_{\mu\nu}$ is the gluon
field strength operator.

Specifically, we would like to extract the matrix elements for the
three lightest glueball states that can be created by dimension-4
gluonic operators.  These are;
\begin{itemize}
\item $0^{++}$ --- Scalar
\begin{equation}
\langle G|{\rm Tr} E^{2}|0 \rangle\;\;\;{\rm and}\;\;\;
\langle G|{\rm Tr} B^{2}|0 \rangle,
\label{eqn:scalar}
\end{equation}
\item $0^{-+}$ --- Pseudoscalar
\begin{equation}
\langle G|{\rm Tr} E \cdot B|0 \rangle,
\end{equation}
\item $2^{++}$ --- Tensor
\begin{equation}
\langle G|{\rm Tr} E_iE_j|0 \rangle\;\;\;{\rm and}\;\;\;
\langle G|{\rm Tr} B_iB_j|0 \rangle.
\label{eqn:tensor}
\end{equation}
\end{itemize}

Previous studies have shown that smooth interpolating fields give
better overlap with the glueball states.  For each lattice irrep of
interest, we compute the correlation matrix for a large basis,
$\{\phi_i\}$, of smeared and fuzzed operators.  For every member of
this basis, the correlation with the appropriate gluon field-strength
operators of Eqns.~(\ref{eqn:scalar}--\ref{eqn:tensor}) is also
computed.  Variational techniques are then employed to find the
coefficients, $c_i$, needed to make an optimal ground-state creation
operator, $\Phi(t)=\sum_i c_i \phi_i(t)$.  The relevant correlators
required to extract both the glueball mass and creation matrix
element are then formed.  These are the smeared-smeared operator
\begin{eqnarray}
\langle 0|\Phi(t)\Phi(0)|0 \rangle - \langle 0|\Phi|0 \rangle^2
\;\;\;\;\;\;\;\;\;\;\;\;\; \nonumber \\
\;\;\;\;\;\; 
\stackrel{t\rightarrow\infty}{-\!\!\!-\!\!\!-\!\!\!-\!\!\!\longrightarrow} 
\langle 0|\Phi|G \rangle \langle G|\Phi|0 \rangle e^{-M_{G}t}
\end{eqnarray}
and the local-smeared operator
\begin{eqnarray}
\langle 0|\Phi(t)F^{2}(0)|0 \rangle -
\langle 0|\Phi|0 \rangle \langle 0|F^{2}|0 \rangle \nonumber \\
\stackrel{t\rightarrow\infty}{-\!\!\!-\!\!\!-\!\!\!-\!\!\!\longrightarrow} 
\langle 0|\Phi|G \rangle \langle G|F^2|0 \rangle e^{-M_{G}t},
\end{eqnarray}
where $|G \rangle$ is the lowest glueball state in the irrep.  The
vacuum subtraction denoted here need only be performed for the
$A_1^{++}$ irrep.  A three-parameter fit to these correlators then
extracts the glueball mass $M_{G}$ and the desired glueball creation
matrix element $\langle G|F^2|0 \rangle$, as well as the glueball
interpolation field matrix element $\langle 0|\Phi|G \rangle$.

\begin{figure}[t]
  \begin{center}
  \leavevmode
  \epsfxsize=\hsize 
  \epsfbox{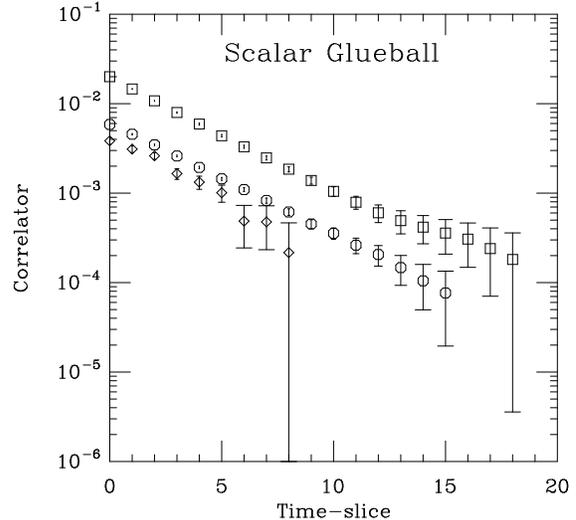}
  \end{center}
  \vspace{-0.5in}
  \caption{$A_1^{++}$ (scalar) correlators.  The
$\langle\Phi|\Phi\rangle$, $\langle {\rm Tr} B^2|\Phi\rangle$ and
$\langle{\rm Tr} E^2|\Phi\rangle$ data are indicated by $\Box$,
$\circ$ and $\Diamond$ respectively.  The $\langle\Phi|\Phi\rangle$
correlator has been rescaled by $2\times 10^{-2}$ for presentation.}
  \label{sc}
\end{figure}

The construction of the appropriate improved gluonic operators on the
lattice proceeds as follows.  For a set of prototype small Wilson
loop shapes (for example, the plaquette, $2\times1$ rectangle, etc.)
a linear combination of the 48 orientations $P_j$ (some will be
degenerate for loops with symmetry) of each prototype, $P$, is taken
which transforms under the appropriate irrep, $R$, of the cubic group
(including parity).  The irreps we need to include are thus the
$A_1^+,A_1^-,E^+$ and $T_2^+$.  The real part of the trace on these
loops is then taken, as all the glueball states of interest are
positive under charge conjugation.  The lattice operators are
\begin{equation}
C^{(R)}_i(P) = \sum_{j=1\dots48} a_{ij}^{(R)} {\rm Re Tr }(1-U(P_j)), 
\end{equation}
where the index $i$ runs over the dimension of the irrep. 
Then, using the standard mapping
\begin{equation}
U_{\mu}(x) = {\cal P}\exp\{-ig\int_x^{x+\hat{\mu}} A_{\mu}(s) \; ds_{\mu}\},
\end{equation}
the operators, $C^{(R)}_i(P)$, are expressed at tree-level in
perturbation theory in terms of the dimension 4 and 6 continuum gluon
field operators transforming under the same irrep of the cubic group.
Once this expansion is performed for all prototypes, a linear
combination of the lattice operators that removes the ${\cal
O}(a^{2}_{s})$ discretization errors is found.  For the scalar and
$E^{++}$ polarizations of the tensor currents, this search is
unnecessary, as conveniently, improving the action generates the
required operators (Tr $E_i^2$ and Tr $B_i^2$).

\section{PRELIMINARY RESULTS}

\begin{figure}[t]
  \begin{center}
  \leavevmode
  \epsfxsize=\hsize 
  \epsfbox{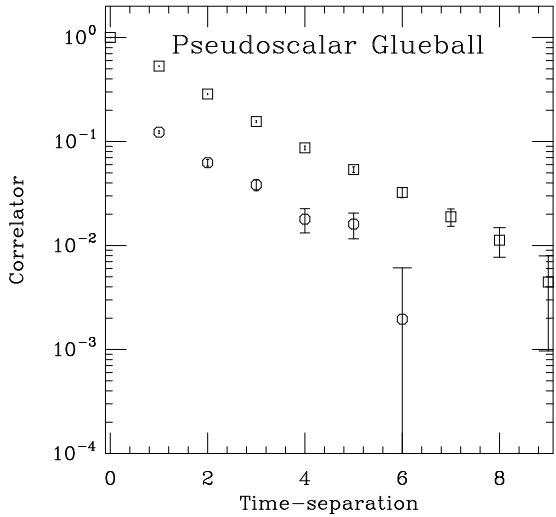}
  \end{center}
  \vspace{-0.5in}
  \caption{$A_1^{-+}$ (pseudoscalar) correlators.  The
$\langle\Phi|\Phi\rangle$ and $\langle{\rm Tr} E\cdot B|\Phi\rangle$
data are indicated by $\Box$ and $\circ$.}
  \label{pssc}
\end{figure}

\begin{figure}[t]
  \begin{center}
  \leavevmode
  \epsfxsize=\hsize 
  \epsfbox{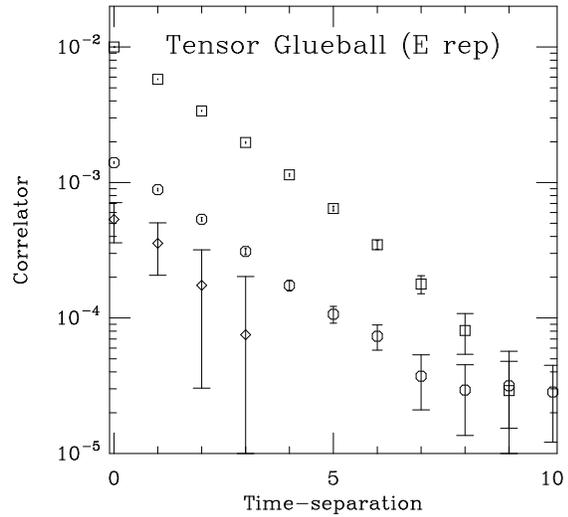}
  \end{center}
  \vspace{-0.5in}
  \caption{$E^{++}$ (tensor) correlators.  The
$\langle\Phi|\Phi\rangle$, $\langle{\rm Tr} B_iB_j|\Phi\rangle$ and
$\langle{\rm Tr} E_iE_j|\Phi\rangle$ data are indicated by $\Box$,
$\circ$ and $\Diamond$ respectively.  The $\langle\Phi|\Phi\rangle$
channel has been rescaled by $10^{-2}$.}
  \label{tensor}
\end{figure}

In Figs.~\ref{sc},~\ref{pssc}~and~\ref{tensor} we present results
from a feasibility study on an $8^3 \times 40$ lattice at $\beta =
2.4$ and anisotropy $a_s/a_t=5$.  These run parameters correspond
\cite {pm1} to a spatial lattice spacing of 0.221(11) fm and temporal
cut-off of 4.5(2) GeV.  For all channels considered, a signal for the
smeared-local correlator can be observed over a sufficiently wide
range of time slices to extract the relevant observables mentioned in
section~\ref{sec:method}.  Correlators with chromoelectric fields are
seen to be significantly more noisy, however.

\section{CONCLUSIONS}

A calculation of the matrix elements relevant for predicting the
branching ratio for radiative decays of $J/\psi$ into glueballs has
been proposed as a useful handle on the phenomenological
identification of the lighter glueball candidates.  This decay should
be a glueball-rich channel, with fairly clear experimental signature.
We have performed exploratory studies of the $0^{++}$, $0^{-+}$ and
$2^{++}$ (in the $E$ irrep) correlator channels, on a lattice with
5:1 anisotropy and with spatial lattice spacing of $a_s\approx0.2$
fm.  They show that anisotropic lattice techniques will allow a
Monte-Carlo calculation of the lattice matrix elements with
reasonable computational resources.

In the future, we will calculate with various lattice sizes and
spacings and take the infinite-volume and continuum limits.  We will
include heavier glueball states and all 20 irreps of $O^{PC}_{h}$.
We will calculate the renormalization constants.

\end{document}